\documentclass{article}

\begin{document}

\title{On $SU(2)$ Yang-Mills Theory and the Maximal Abelian Gauge}
\author{O. Oliveira \\
        orlando@teor.fis.uc.pt \\
        Centro de F\'{\i}sica Computacional \\
        Departamento de F\'{\i}sica \\
        Universidade de Coimbra \\
        3004-516 Coimbra \\
        Portugal }

\maketitle

\begin{abstract}
The problem of identifying the dynamical degrees of freedom for $SU(2)$
gauge theory is discussed. After studying $SU(N)$ theories, 
it is shown that classical pure $SU(2)$ gauge theory is equivalent to an
abelian theory. Finally, we prove that in the Euclidean formulation, the 
Maximal Abelian Gauge correctly identifies the abelian degrees of freedom 
of the $SU(2)$ theory.
\end{abstract}


Gauge field theories play a major role in High Energy Physics. The dynamics
of gauge theories has been investigated mainly with perturbative methods. 
However, not all 
physical phenomena can be explained within perturbation theory. 
For example, the understanding of quark confinement in hadrons
requires non-perturbative techniques applied to Quantum Chromodynamics (QCD).
To explore the non-perturbative regimes one would like to 
identify the relevant dynamical degrees of freedom and to solve the field
equations. For QCD the gluon degrees of freedom are, typically, described as
copies of the QED photon field. This language seems to be useful to 
understand the high energy limit of QCD, but not for its
low energy regime.

Recently, a number of authors \cite{FaNi98,Cho99,Sha} addressed the problem 
of the relevant degrees of freedom for $SU(2)$ Yang-Mills theory. 
A  parameterization of the gauge fields was proposed and effective actions
in terms of scalar fields derived. Furthermore, a number of 
results have been extended to $SU(N)$ theories \cite{FaNi98a,FaNi99,Li99}.

For $SU(2)$, in \cite{FaNi98} Faddeev and Niemi proposed a Skyrme like 
effective action to describe the low energy limit, written in terms of what 
they assumed to be the relevant degrees of freedom. 
Their work was inspired on a suggestion from 'tHooft and Polyakov 
\cite{tHPo,Poly}, 
that tried to describe color confinement as a dual Meissner effect, and on a 
generalization of the Wu-Yang ansatz \cite{WuYa}. Along the same lines,
in \cite{FaNi98a} they parameterized the gluon 
fields\footnote{See also \cite{FaNi99}.} for an $SU(N)$ gauge theory and wrote
a generalized Skyrme effective action for $SU(N)$ Yang-Mills theory. 

In \cite{Cho99,Cho80} Cho wrote a parameterization of the 
gauge fields for $SU(2)$ in terms of a covariant constant scalar field in the 
adjoint representation and a photon-like field. Then, he proved 
abelian dominance for Wilson loops.
Meanwhile, and within the philosophy of the work of Faddeev and Niemi, 
in \cite{ChoLeePak} another generalized Skyrme-Faddeev action was derived as 
an effective action for the low energy limit of $SU(2)$ gauge theory.

For $SU(N)$ theories, Li \textit{et al.} \cite{Li99} used the technique
suggested in \cite{Cho99} to write a general parameterization for the gluon 
fields from which they built an effective action.

So far the main work has been centered on building effective actions 
describing the low energy limit of $SU(N)$ gauge theories. Presently, a
number of slightly different effective actions has been derived. It would
be interesting to investigate how they are related to each other. In this 
paper we do not seek to discuss effective actions. Instead we 
look for possible ways of identifying the dynamical degrees of freedom via 
gauge fixing of the original theory.

The paper is organized as follows. We use the procedure of
\cite{Cho99} to write the non-abelian field $A^a_\mu$ for $SU(N)$ theories. 
In particular, for $SU(2)$ we built a general parameterization of the gauge 
field. It follows that the dynamics of classical pure $SU(2)$ maps the
dynamics of an abelian theory. Finally, we prove that, in the Euclidean 
formulation, the Maximal Abelian Gauge reduces the gluon field $A^a_\mu$ to
the photon field of the hidden abelian theory and comment on possible 
implications for lattice simulations.

\section{Gluon Fields for $SU(N)$ Gauge Theories}

The lagrangian for $SU(N)$ gauge theories reads
\begin{equation}
 \mathcal{L} \, = \, - \frac{1}{4} \, F^a_{\mu\nu} \, F^{a \, \mu\nu}
\end{equation}
where
\begin{equation}
 F^a_{\mu\nu} \, = \, \partial_\mu A^a_\nu \, - \,
                      \partial_\nu A^a_\mu  \, - \,
                      g f_{abc} A^b_\mu A^c_\nu
 \mbox{ };
\end{equation}
$A^a_\mu$ are the gluon fields. 

Let $n^a$ be a covariant constant real scalar field in the adjoint 
representation. From the definition it follows that
\begin{equation}
 D_\mu \, n ^a \, = \, 
    \partial_\mu n^a \, + \, i g \left( F^b \right)_{ac} A^b_\mu n^c
  \, = \, 0 \, ; \label{Dn}
\end{equation}
the generators of the adjoint representation are defined in the
usual way $\left( F^b \right)_{ac} \, = \, -i \, f_{bac} \,$. 
Given a gluon field it is always possible to solve the above equations for 
the scalar field $n^a$. In this way it is possible to define a map from the 
gluon field to $n$. Let us reverse the argument. Multiplying (\ref{Dn}) by 
$\left( F^d \right)_{ea} n^d$ and solving the equations for the gauge 
fields we get, after some algebra,
\begin{eqnarray}
 & &   \left( F^d \right)_{ea} \, n^d \, \partial_\mu n^a \, + \,
 i g \, ( n \cdot A_\mu ) \, n^e \, - \,
  i g \, ( n \cdot n ) A^e_\mu \, + \,
\nonumber \\
 & &
 \left[  \left( D^b D^e \right)_{dc} \, n^d \, n^c \, - \,
         \left( D^d D^c \right)_{ed} \, n^d \, n^c \right] A^b_\mu
 \, = \, 0 \, ,
 \label{eqn1}
\end{eqnarray}
where
\begin{equation}
\left( D^a \right)_{bc} \, = \, d_{abc} \, = \,
   \frac{1}{4}
   \mbox{Tr} \left( \lambda^a \, \left\{ \,\lambda^b \, , \,
                                           \lambda^c \, \right\}
             \right)
\end{equation}
and $\lambda^a$ are the Gell-Mann matrices.
Writing the gauge fields as
\begin{equation}
   A^a_\mu \, = \, \hat{A}_\mu n^a \, + \, X^a_\mu \, ,
\end{equation}
equations (\ref{eqn1}) become
\begin{eqnarray}
 &  &  \left( F^d \right)_{ea} \, n^d \, \partial_\mu n^a \, + \, 
 i g \, ( n \cdot X_\mu ) \, n^e \, - \,
  i g \, ( n \cdot n ) X^e_\mu \, + \, 
\nonumber \\
 & &
 \left(  d_{dba} \, d_{ace} \, - \,
         d_{dca} \, d_{aeb} \right) n^d \, n^c X^b_\mu
 \, = \, 0 \, . \label{eqn2}
\end{eqnarray}
By a convenient definition of $\hat{A}_\mu$ it follows, without loss
of generality, that
\begin{eqnarray}
  &  &  X^a_\mu \, = \, \frac{1}{ig} \, \left( F^d \right)_{ea} \,
                   \frac{ n^d \, \partial_\mu n^a }{ n \cdot n } \, + \,
                 Y^a_\mu  \\
  &  &  n \cdot Y_\mu \, = \, 0. 
\end{eqnarray}

In terms of $\hat{A}_\mu$, $n^a$ and $Y^a_\mu$, the gauge fields are given by
\begin{equation}
 A^a_\mu \, = \, \hat{A}_\mu n^a \, + \,
                 \frac{1}{ig} \, \left( F^c \right)_{ab} \,
                              \frac{n^c \, \partial_\mu n^b}{n \cdot n} \, + \,
                 Y^a_\mu
 \, ; \label{A}
\end{equation}
with $n$ and $Y$ verifying the constraints
\begin{eqnarray}
 & & n \cdot Y_\mu \, = \, 0 \, ,  \label{nY} \\
 & & D_\mu n^a \, = \, 0  . \label{Dn1}
\end{eqnarray}
If (\ref{A}) together with (\ref{nY}) and (\ref{Dn1}) provide a complete
parameterization of the gluon fields, the total number of independent fields
on both sides of (\ref{A}) should be the same.
On the l.h.s., the number of degrees of freedom is $2 (N^2 - 1)$.
For the r.h.s., depending on how you perform the 
counting\footnote{At this point one should alert the reader that the fields 
$n$, $\hat{A}_\mu$ and $Y^a_\mu$ should not be regarded as free fields and 
it is not clear that the usual naive counting provides the correct answer.} 
it can be made as large as $2 (N^2 - 1)$. 
Let us consider $SU(2)$ for simplicity. 
For the r.h.s., the number of d.o.f being 3 
from $n$, 2 (3) for a massless (massive) $\hat{A}_\mu$ and possible additional
d.o.f. coming from $Y^a_\mu$. Despite the constraints, easily one arrives
at a number of 6 independent fields. 

Now, let us look at the gauge transformation properties of $n$, $Y$ and 
$\hat{A}$. Since $n$ is covariant constant, it follows that 
$ -i \, \left( F^c \right)_{ab} \, n^c \, \partial_\mu n^b \, / 
(n \cdot n)$ belongs to the adjoint representation of the gauge group. 
Demanding that $Y$ is also in the adjoint representation, then
\begin{equation}
   \hat{A}_\mu \, \longrightarrow \, \hat{A}_\mu \, + \,
                                     \frac{1}{g} \, 
                                         \frac{n \cdot \partial_\mu \omega}
                                              {n \cdot n} \, .
\end{equation}
Constraints (\ref{nY}) and (\ref{Dn1}) are scalars under the gauge group and
the parameterization (\ref{A}) provides a complete gauge invariant 
decomposition of the gluon field $A^a_\mu$.

For $SU(N)$ gauge theories it follows from (\ref{Dn}) and (\ref{A}) that
\begin{equation}
    n \partial_\mu n \, = \, \frac{1}{2} \, \partial_\mu n^2 \, = \, 0,
   \label{t1}
\end{equation}
and one can always choose $n^2 \, = \, 1$.

\section{$SU(2)$ Gauge Theory}

In this section we are going to study the implications of
(\ref{A}), (\ref{nY}) and (\ref{Dn1}) for
$SU(2)$ gauge theory. The scalar field $n$ can be
parametrized by the functions $\theta_1$ and $\theta_2$ as
\begin{equation}
   n \, = \, \left( \begin{array}{c}
                     \sin \theta_1 \, \cos \theta_2  \\
                     \sin \theta_1 \, \sin \theta_2  \\
                     \cos \theta_1
                    \end{array}
             \right) \, .
   \label{nn}
\end{equation}
The field $Y$ is orthogonal to $n$ and can be parametrized in terms of scalar
fields belonging to the adjoint representation of the gauge group as 
(see the appendix for definitions)
\begin{equation}
  Y^a_\mu \, = \, B_\mu m^a \, + \, C_\mu p^a,
\end{equation}
where $B_\mu$ and $C_\mu$ are gauge invariant vector fields. The constraint
(\ref{Dn}) requires
\begin{equation}
 B_\mu \, = \, C_\mu \, = \, 0 \, ;
\end{equation}
note that the second term in (\ref{A}) generates components along $p$ and 
$m$ directions. Then, the gluon field becomes
\begin{eqnarray}
  A^1_\mu & = &
     \hat{A}_\mu \, \sin \theta_1 \, \cos \theta_2 \, - \,
           \frac{1}{g} \, \sin \theta_1 \, \cos \theta_1 \, \cos \theta_2
                          \, \partial_\mu \theta_2 \, - \,
            \frac{1}{g} \, \sin \theta_2 \, \partial_\mu \theta_1 \, , 
  \nonumber \\
  A^2_\mu & = &
      \hat{A}_\mu \, \sin \theta_1 \, \sin \theta_2 \, - \,
            \frac{1}{g} \, \sin \theta_1 \, \cos \theta_1 \, \sin \theta_2
                        \, \partial_\mu \theta_2 \, + \,
            \frac{1}{g} \, \cos \theta_2 \, \partial_\mu \theta_1 \, , 
  \nonumber \\
  A^3_\mu & = &
      \hat{A}_\mu \, \cos \theta_1 \, + \,
      \frac{1}{g} \, \sin^2 \theta_1
                  \, \partial_\mu \theta_2 \, ,
\end{eqnarray}
the gluon field tensor
\begin{equation}
   F^a_{\mu\nu} \, = \, n^a \, \mathcal{F}_{\mu\nu}
\end{equation}
with
\begin{equation}
 \mathcal{F}_{\mu\nu} \, = \, \partial_\mu \mathcal{A}_\nu \, - \, 
                              \partial_\nu \mathcal{A}_\mu
\end{equation}
and
\begin{equation}
 \mathcal{A}_\mu \, = \, \hat{A}_\mu \, - \, \frac {1}{g} \cos \theta_1 
             \left( \partial_\mu \theta_2  \right).
 \label{abelian}
\end{equation}
The action is
\begin{equation}
  S \, = \, - \frac{1}{4} \, \int d^4x \, \mathcal{F}^2
\end{equation}
and the classical equations of motion
\begin{equation}
  n^a \, \partial_\nu \mathcal{F}^{\mu\nu} \, = \, 0 \, .
  \label{eqm}
\end{equation}
Ignoring the trivial solution, (\ref{eqm}) shows that classical pure 
$SU(2)$ is equivalent to an abelian theory, with the abelian field given 
by (\ref{abelian}). For the quantum theory, due to the nonlinear relation
between the original gauge fields and $\mathcal{A}$, 
\begin{eqnarray}
  A^1_\mu & = &
     \mathcal{A}_\mu \, \sin \theta_1 \, \cos \theta_2 \, - \,
            \frac{1}{g} \, \sin \theta_2 \, \partial_\mu \theta_1 \, , 
  \nonumber \\
  A^2_\mu & = &
      \mathcal{A}_\mu \, \sin \theta_1 \, \sin \theta_2 \, + \,
            \frac{1}{g} \, \cos \theta_2 \, \partial_\mu \theta_1 \, , 
  \nonumber \\
  A^3_\mu & = &
      \mathcal{A}_\mu \, \cos \theta_1 \, + \,
      \frac{1}{g}  \, \partial_\mu \theta_2  \, ,
  \label{geral}
\end{eqnarray}
the relation is not so simple.

If classical $SU(2)$ Yang-Mills theory is equivalent to an abelian theory, 
is it possible to keep only the abelian field by choosing a convenient
gauge?  The question can be answered positively in the Euclidean formulation.

From now on we will assume to be in the Euclidean formulation.
In the maximal abelian gauge (MAG) we look for fields which maximize
\begin{eqnarray}
  R[A] & = &  - \int d^4x \, \left[  \, \left( A^1_\mu \right)^2 \, + \,
                              \left( A^2_\mu \right)^2 \, \right] 
  \nonumber \\
       & = &
         - \int d^4x  \, \left[  \, \left( \mathcal{A}_\mu \right)^2 \,  
                                   \sin^2 \theta_1  \, \sin^2 \theta_2 
       \, + \,
                              \frac{1}{g} \left( \partial \theta_1 \right)^2
                         \, \right]
  \label{MAG}
\end{eqnarray}
along each gauge orbit. The absolute maximum of (\ref{MAG}) is realized by
two classes of fields :
i) $\mathcal{A} \, = \, \partial^2 \theta_1 \, = \, 0$ (type I) and
ii) $\sin^2 \theta_1 \, = \, 0$ (type II), both having 
$A^1_\mu \, = \, A^2_\mu \, = \, 0$. The solutions of type I are 
vacuum solutions ($F = 0$) and can be gauged to the null solution. 
For type II solutions the gluon field is
\begin{equation}
 A^3_\mu \, = \, \pm \mathcal{A}_\mu \, + \, \frac{1}{g} \, 
       \partial_\mu \theta_2
\end{equation}
and by a gauge transformation \cite{Ga} can be reduced (up to a sign) to the 
abelian field $\mathcal{A}$. Note that the type II solution is essentially
the $\hat{A}_\mu$ field,
\begin{equation}
 A^3_\mu \, = \,  \left\{
       \begin{array}{l}
         \hat{A}_\mu \, + \, \frac{2}{g} \partial_\mu \theta_2 \, , \\
          \pm \hat{A}_\mu \, , \\
          -\hat{A}_\mu \, + \, \frac{2}{g} \partial_\mu \theta_2 \, .
       \end{array}
                 \right.
\end{equation}

For what concerns the quantum theory, the jacobian relating $A^a_\mu$ and 
$\mathcal{A}_\mu$ is now unitary, 
i.e., in the Maximal Abelian Gauge, pure $SU(2)$ 
Yang-Mills theory is equivalent to an abelian theory. 

The inclusion of fermions changes the relation between the two theories. 
In terms of the abelian field $\mathcal{A}$ the fermionic 
covariant derivative in the MAG is
\begin{equation}
 \left( D^a_\mu \right) \, = \, 
   \left( \begin{array}{c}
                  \partial_\mu  \\
                  \partial_\mu  \\
                  \partial_\mu \, + \, \frac{ig}{2} \sigma^3 \mathcal{A}
          \end{array}
   \right) \, . 
  \label{dfermion}
\end{equation}
$SU(2)$ gauge theory is formally identical to strong coupled QED with two 
opposite charges per quark flavor. In principle, the quest for solutions of
the classical theory looks much simpler in the MAG compared to the original 
formulation. Therefore, MAG opens a window to investigate the relevance of
the different classical configurations to the characteristics of the quark 
dynamics \cite{OO}.

The above results for the MAG can be combined with lattice simulations
to test abelian dominance and, hopefully, understand the role of 
Gribov copies on different observables. Our analysis shows that 
condition (\ref{MAG}) does not remove completely the gauge ambiguity. 
We have used the remaining freedom to gauge $A^3_\mu$ to the abelian field. 
However, even after doing that we are left with two solutions 
$A^3_\mu \, = \, \pm \mathcal{A}_\mu$ per space-time point, meaning that the 
number of Gribov copies is infinite. 

On the lattice the situation is similar. If we are close to the continuum, 
one can write the link variables as
\begin{equation}
  \mathcal{U}_\mu (x) \, = \, \exp \left\{ \frac{i}{2} g a \mathcal{A}_\mu
                                           \sigma^3 \right\}
  \, = \,
  \left(  \begin{array}{cc}
          e^{ \pm i \mathcal{A}_\mu }  &  0  \\
           0    &    e^{ \mp i \mathcal{A}_\mu }
          \end{array}
  \right) \, ,
  \label{link}
\end{equation}
i.e. the number of Gribov grows with lattice volume.
However, knowing that the link should look like (\ref{link}), one can
test a given gauge fixing algorithm to check if it identifies
correctly the absolute maximum of (\ref{MAG}) and in this way
study the influence of Gribov copies on the Green's functions of our theory
\cite{Stack}.

\section{Results and Conclusions}

In this paper we argued that it is possible to write the gluon fields 
for $SU(N)$ gauge theories in terms of scalar and vector fields. Carefully
chosen, these fields can simplify the classical equations of motion.

For $SU(2)$ Yang-Mills theory our choice of the parameterization shows
that the classical theory is equivalent to an abelian theory. 
Furthermore, it was proved that in the MAG (Euclidean formulation) 
the quantum theory is equivalent to an abelian theory. This
result shows abelian dominance for the $SU(2)$ theory. In what concerns
Gribov copies, we found that in the MAG the number of copies is
infinite. This is true also for lattice simulations.
For lattice calculations we suggest a test to check
if gauge fixing algorithms correctly identify the absolute maximum of
the lattice version of (\ref{MAG}).

Despite additional technical complications, our discussion for $SU(2)$ can 
be extended to $SU(3)$ \cite{OO3} and other higher rank special unitary 
groups. From our study, naively, one expects to observe richer structure 
as the rank of the group is increased. Currently, we are 
involved in extending this work to QCD.

\section*{Appendix}

In three dimensions one can define the following orthogonal unitary vectors
\begin{equation}
   n \, = \, \left( \begin{array}{c}
                     \sin \theta_1 \, \cos \theta_2  \\
                     \sin \theta_1 \, \sin \theta_2  \\
                     \cos \theta_1
                    \end{array}
             \right) \, , \mbox{    }
   m \, = \, \left( \begin{array}{c}
                     \cos \theta_1 \, \cos \theta_2  \\
                     \cos \theta_1 \, \sin \theta_2  \\
                     -\sin \theta_1
                    \end{array}
             \right) \, ,  \mbox{    }
   p \, = \, \left( \begin{array}{c}
                     \sin \theta_2  \\
                     - \cos \theta_2  \\
                       0
                    \end{array}
             \right) \, .
\end{equation}
The unitary vectors $n$, $m$ and $p$ verify the relations
\begin{eqnarray}
 \partial_\mu n & = & m \left( \partial_\mu \theta_1 \right) \, - \, 
                        p \sin \theta_1 
                          \left( \partial_\mu \theta_2 \right) \, ,
 \\
 \partial_\mu m & = & - \, n \left( \partial_\mu \theta_1 \right) \, - \, 
                        p \cos \theta_1
                          \left( \partial_\mu \theta_2 \right) \, ,
 \\
 \partial_\mu p & = & \left( n \sin \theta_1 \, + \, 
                              m \cos \theta_1 \right) \, 
                       \left( \partial_\mu \theta_2 \right)
\end{eqnarray}
and
\begin{equation}
 \epsilon_{abc} \, n^b \, p^c \, = \, m^a \,\,\,\,\,\,
 \mbox{         and circular permutations.}
\end{equation}


\end{document}